\documentclass[
    ,final            
  ]
  {aipproc}

\layoutstyle{6x9}

\newcommand{\teff}{\mbox{${T}_{\rm eff}$}}

\newcommand{\msol}{\mbox{${\rm M}_{\odot}$}}

\newcommand{\lsol}{\mbox{${\rm L}_{\odot}$}}

\newcommand{\simgt}{\lower.5ex\hbox{$\; \buildrel > \over \sim \;$}}
\newcommand{\simlt}{\lower.5ex\hbox{$\; \buildrel < \over \sim \;$}}
\begin{document}

\title{The effect of using different EOS in modelling the $\alpha$ \emph{Centauri} binary system.}


\author{A. Miglio}{
  address={Institut d'Astrophysique et G\'eophysique de l'Universit\'e de Li\`ege, Belgium}
}

\begin{abstract}
In this preliminary study we investigate the effects of using different equations of state (CEFF and OPAL) in the calibration of the binary system $\alpha$ \emph{Centauri}. Constraints coming from the detection of acoustic oscillations in $\alpha$ \emph{Centauri} A and B are included in the modelling.
\end{abstract}

\maketitle


\section{Introduction}
The visual binary system $\alpha$ \emph{Centauri} represents a promising target to test our understanding of stellar structure and evolution due to its numerous and stringent observational constraints, including the recent detection of acoustic oscillations frequencies in both components of the system \cite{bouchy}, \cite{carrier}.

\section{Modelling $\alpha$ Centauri}
The calibration of the system consists in defining a goodness-of-fit measurement ($\chi^2$), that includes both seismic and non-seismic constraints weighted with their observational uncertainties, and then minimizing the $\chi^2$ using a Levenberg-Marquardt optimization procedure. Models for components A and B, computed with the same initial chemical composition, are fitted to their observational constraints at the same age. We take as observational constraints the small ($\delta\nu$) and large ($\Delta\nu$) frequency separations (see e.g. \cite{jcd_sep}) derived from the observed oscillation frequencies by \cite{bouchy} and \cite{carrier} as well as the effective temperature, luminosity and metallicity (Table \ref{tab:obs1}). A more detailed description of the calibration procedure and of the choice of observational constraints will be reported in \cite{inprep}.
\begin{center}
\begin{figure}
  \includegraphics[angle=-90,width=.5\textwidth]{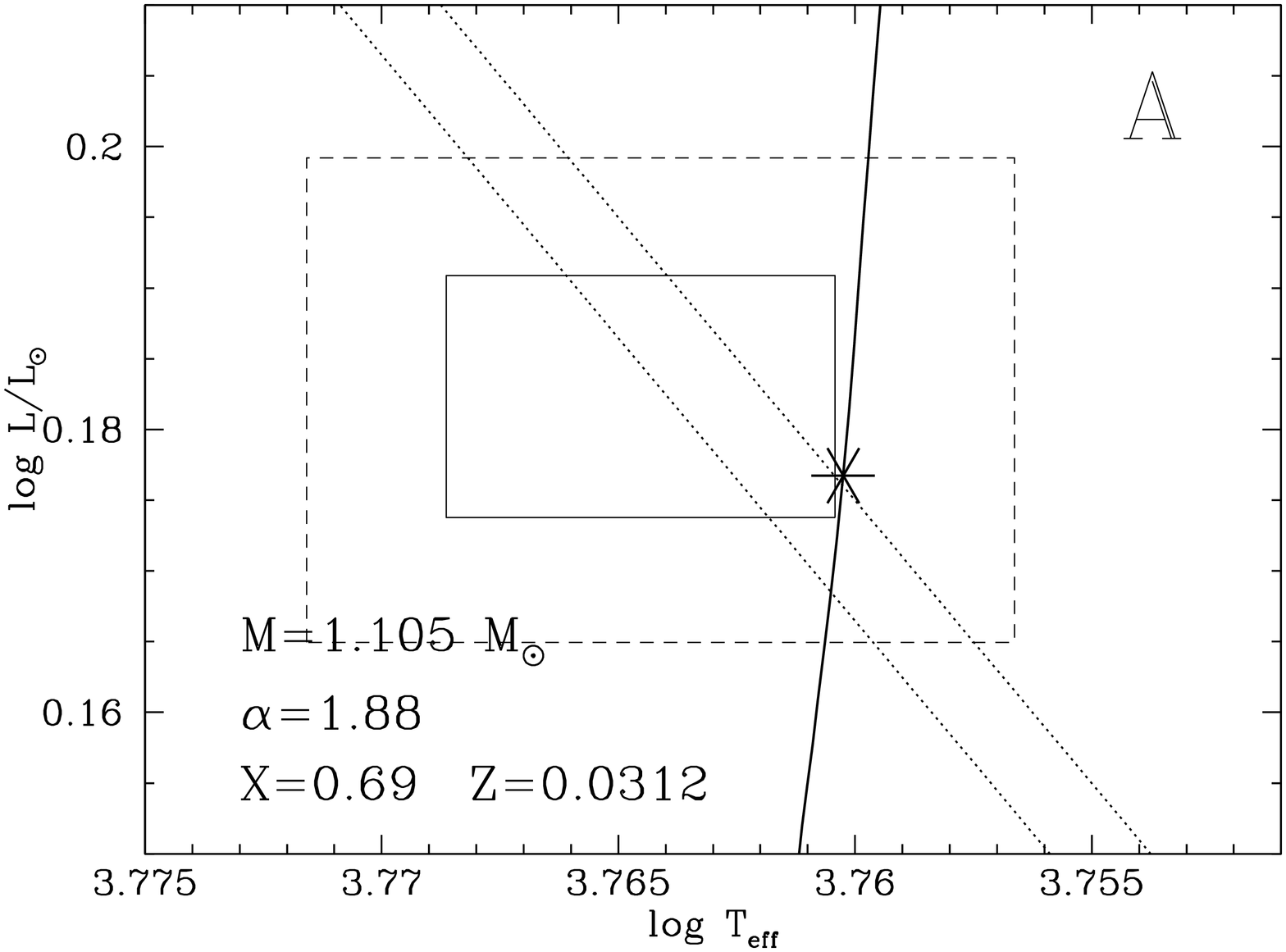}
  \includegraphics[angle=-90,width=.5\textwidth]{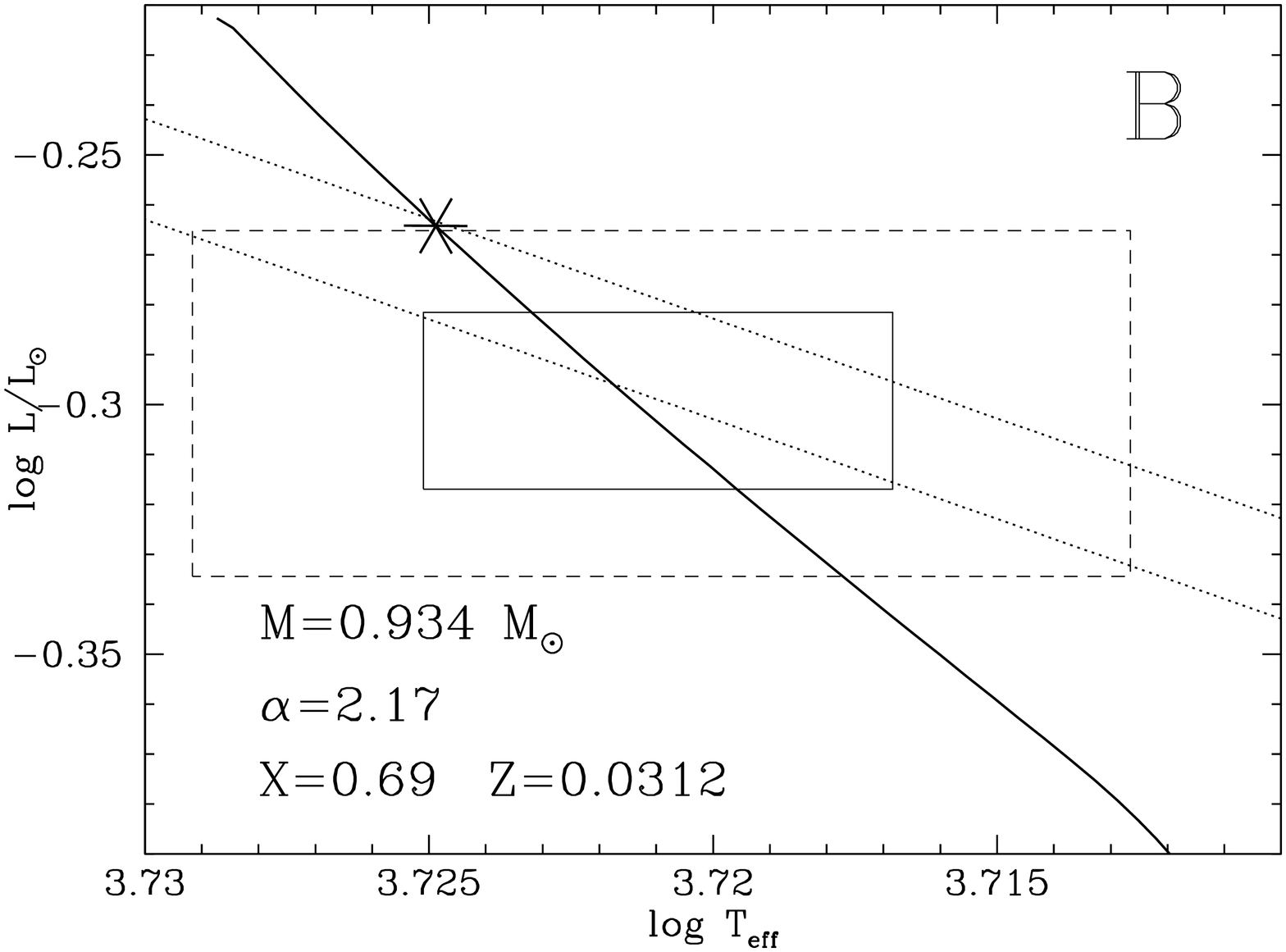}
 \caption{Position of models A0 and B0 in the HR diagram with 1$\sigma$ and  2$\sigma$ error boxes in  luminosity, effective temperature and radius (2$\sigma$, assuming the values determined by \cite{kervella}).}\label{fig:hrAB}
\end{figure}
\end{center}
\begin{center}
\begin{figure}
  \includegraphics[angle=-90,width=.5\textwidth]{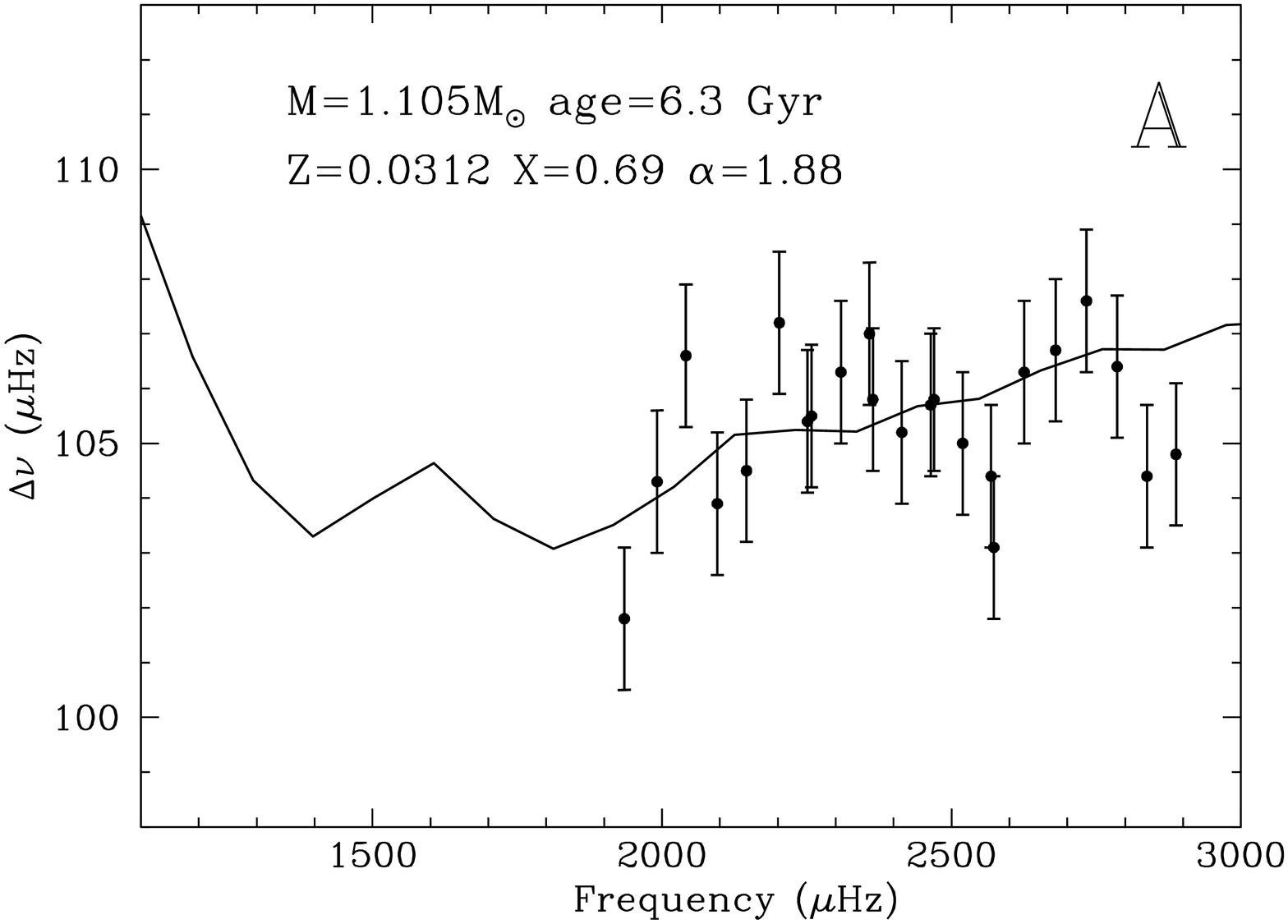}
  \includegraphics[angle=-90,width=.5\textwidth]{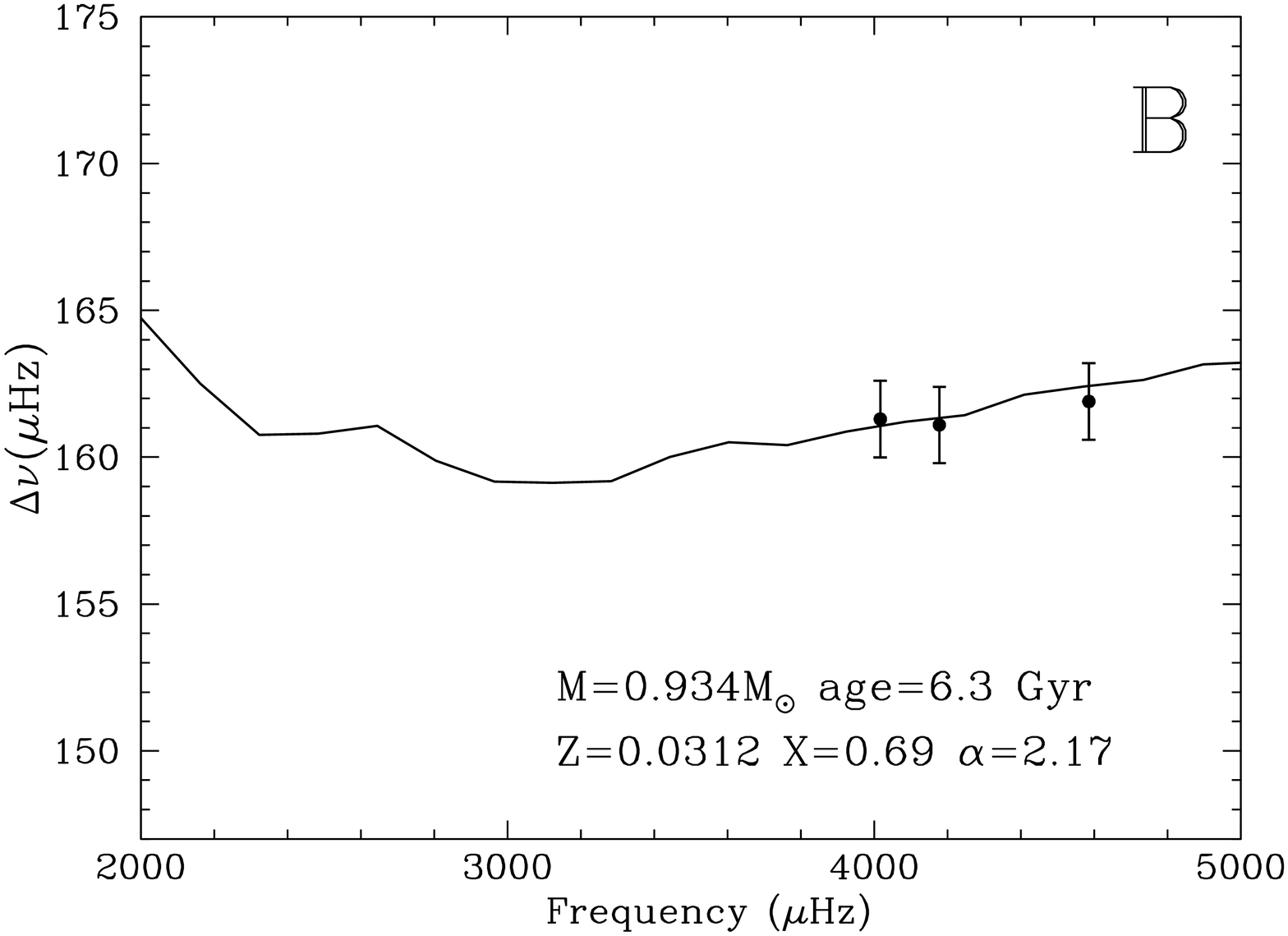} 
\caption{Comparison between observed and predicted (model A0 and B0) large frequency separation $\Delta\nu$ for $\alpha$ Cen A and B. }\label{fig:lsAB}
\end{figure} 
\end{center}
We take as free parameters the age, the initial chemical composition (X,Z) and the mixing length parameter for each component: $\alpha_A$ and $\alpha_B$. The masses are fixed and taken equal to the values given by \cite{pourbaix} ($\rm M_{\rm A}$= 1.105 \msol~and $\rm M_{\rm B}$=0.934 \msol). We have also allowed a variation of the masses within their error bars \cite{inprep}, but its effect is not relevant while studying the effect of different EOS on the calibration of the system.

The stellar models are computed using CLES (Code Li\'egeois d'Evolution  Stellaire). The opacity tables are those of OPAL96 \cite{iglesias96} complemented at $T < 6000$ with Alexander and Ferguson opacities \cite{alexander}. The solar Z-distribution from \cite{grevesse} is adopted in the opacities and in the equation of state.
Nuclear reaction rates are taken from \cite{caughlan} and the screening factor from \cite{salpeter}. Convective zones are treated with the classical mixing length theory \cite{bohm} with the formulation by \cite{cox};  atmospheric boundary conditions \cite{kurucz98} are applied at $T=T_{\rm eff}$.
The code includes microscopic diffusion of H,He and Z using the subroutine by \cite{thoul94}. The equation of state can be chosen among CEFF \cite{jcdceff} and OPAL \cite{rogers02}.

\begin{center}
\begin{figure} 
\includegraphics[angle=-90,width=.5\textwidth]{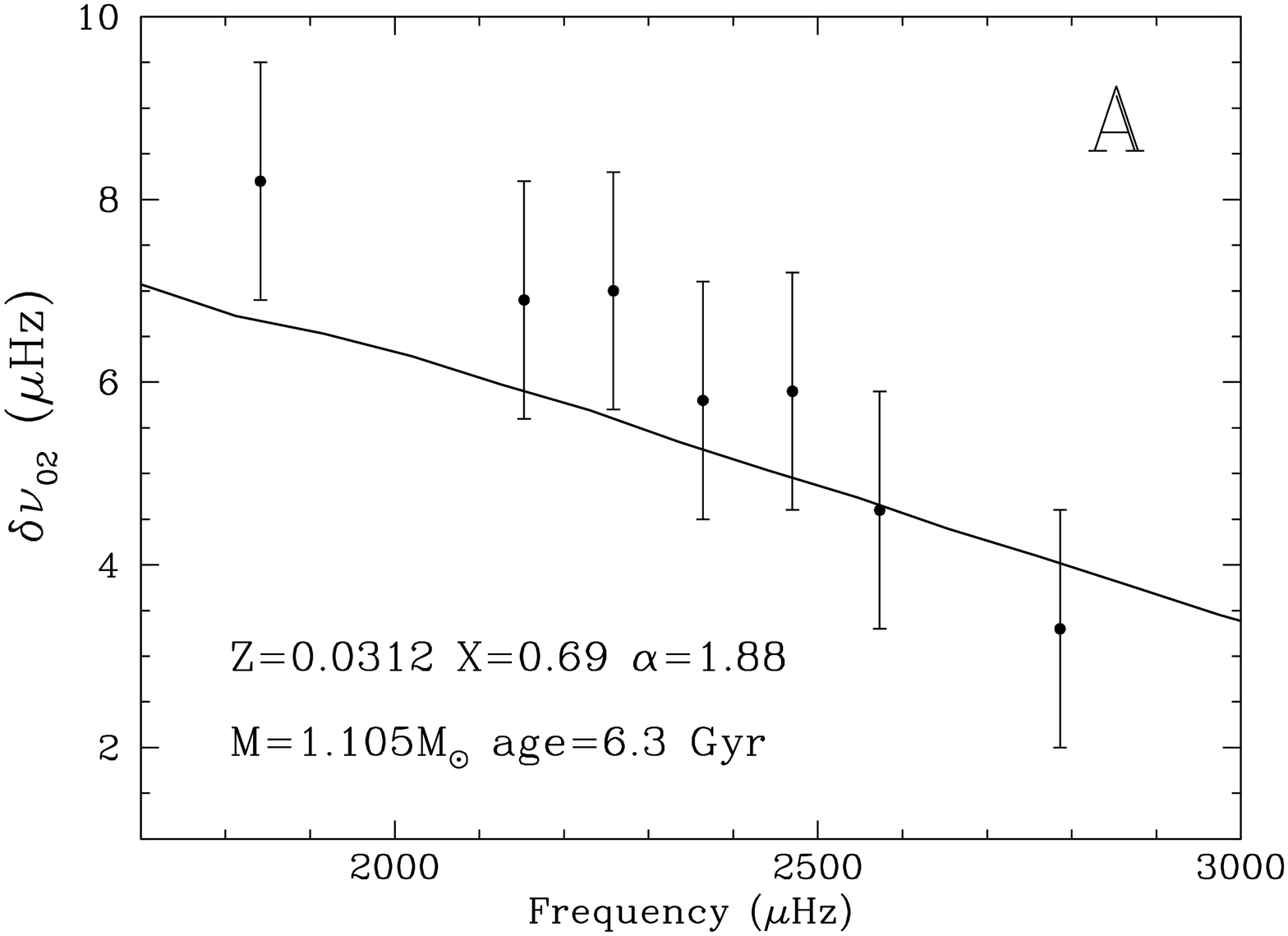}
  \includegraphics[angle=-90,width=.5\textwidth]{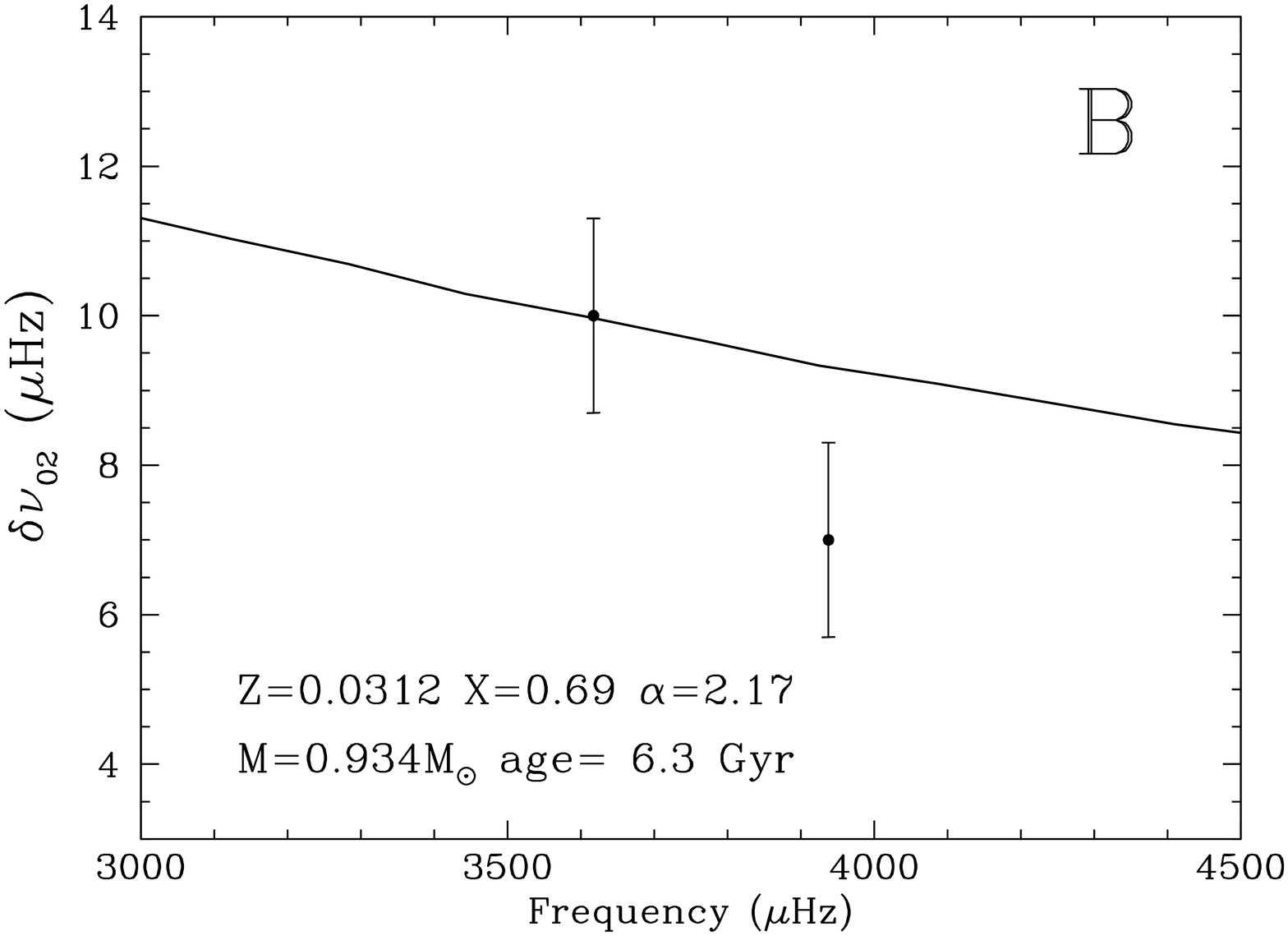}
 \caption{Observed versus predicted small frequency separation ($\delta\nu$).}\label{fig:ssAB}
\end{figure}
\end{center}

Fig. \ref{fig:hrAB} shows the HR diagram location of the $\alpha$ Cen A and B models that best fit the classical (\teff, L) and seismic observables (hereafter A0 and B0, see Table \ref{tab:bestmod}); in Figure \ref{fig:lsAB} and \ref{fig:ssAB} we plot the corresponding large and small separations.
\begin{table}
\begin{tabular}{cccc}
  \hline
  & A & B & Ref \\
\hline
$T_{\rm eff}$ (K) & 5810 $\pm$ 50 & 5260  $\pm$ 50&  \cite{magain97} \\
L/\lsol & 1.522$\pm$0.030 & 0.503$\pm$0.020& \cite{eggenberger04} \\
$(Z/X)_{\rm S}$ & 0.039$\pm$0.006 &0.039$\pm$0.006& \cite{thoul03}\\ 
\hline
 \end{tabular}
\caption{\small Non-asteroseismic constraints assumed in the calibration}\label{tab:obs1}
\end{table}

\section{Intrinsic difference} 
\begin{figure}
  \includegraphics[angle=-90,width=.6\textwidth]{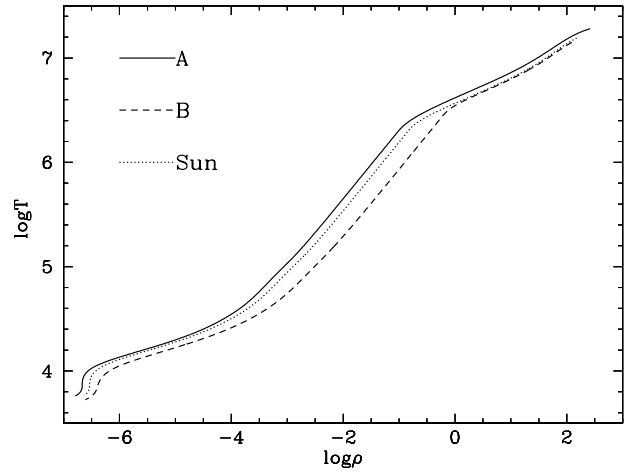}
  \caption{Internal temperature-density profile of reference models A0 and B0 compared to the profile of a solar model (dotted line).}\label{fig:struct}
\end{figure}

In order to analyze the differences between CEFF and OPAL equations of state, we take the internal structure ($T-\rho$) of the best models (A0 and B0) calibrated using OPAL and we estimate $c^2$ and $\Gamma_1$ using CEFF. In Fig. \ref{fig:struct} we show the internal structure of the models A0 and B0 as well as the structure of a solar model for comparison.

The differences between sound speed and first adiabatic exponent, due solely to the use of a different EOS, are shown in  Fig. \ref{fig:dgamma} and \ref{fig:dc2}. These differences are of the same order of those predicted in solar models \cite{basu}, they appear to be larger in the lower-mass model B0 than in model A0 and are mainly located in the hydrogen and helium ionization regions.
\begin{table}
\begin{tabular}{cccccc}
  \hline
Model  & M/\msol & $\alpha$ & Age (Gyr) & $X$ & $Z$ \\
\hline
A0 & 1.105 & 1.88 & 6.3 & 0.690 & 0.0312 \\ 
B0 & 0.934 & 2.17 & 6.3 & 0.690 & 0.0312 \\ 
\hline
 \end{tabular}
\caption{\small Model parameters for reference models A0 and B0.}\label{tab:bestmod}
\end{table}
Such a small difference in the internal structure of a model propagates in a variation of the observables of each model (e.g. effective temperature, luminosity) that could be easily compensated by a re-adjustment of the free parameters in our modelling.  Slightly changing the initial chemical composition, the mixing length parameter or the age, would easily let the models computed with different equations of state satisfy the observational constraints. 
\begin{center}
\begin{figure}
  \includegraphics[angle=-90,width=.5\textwidth]{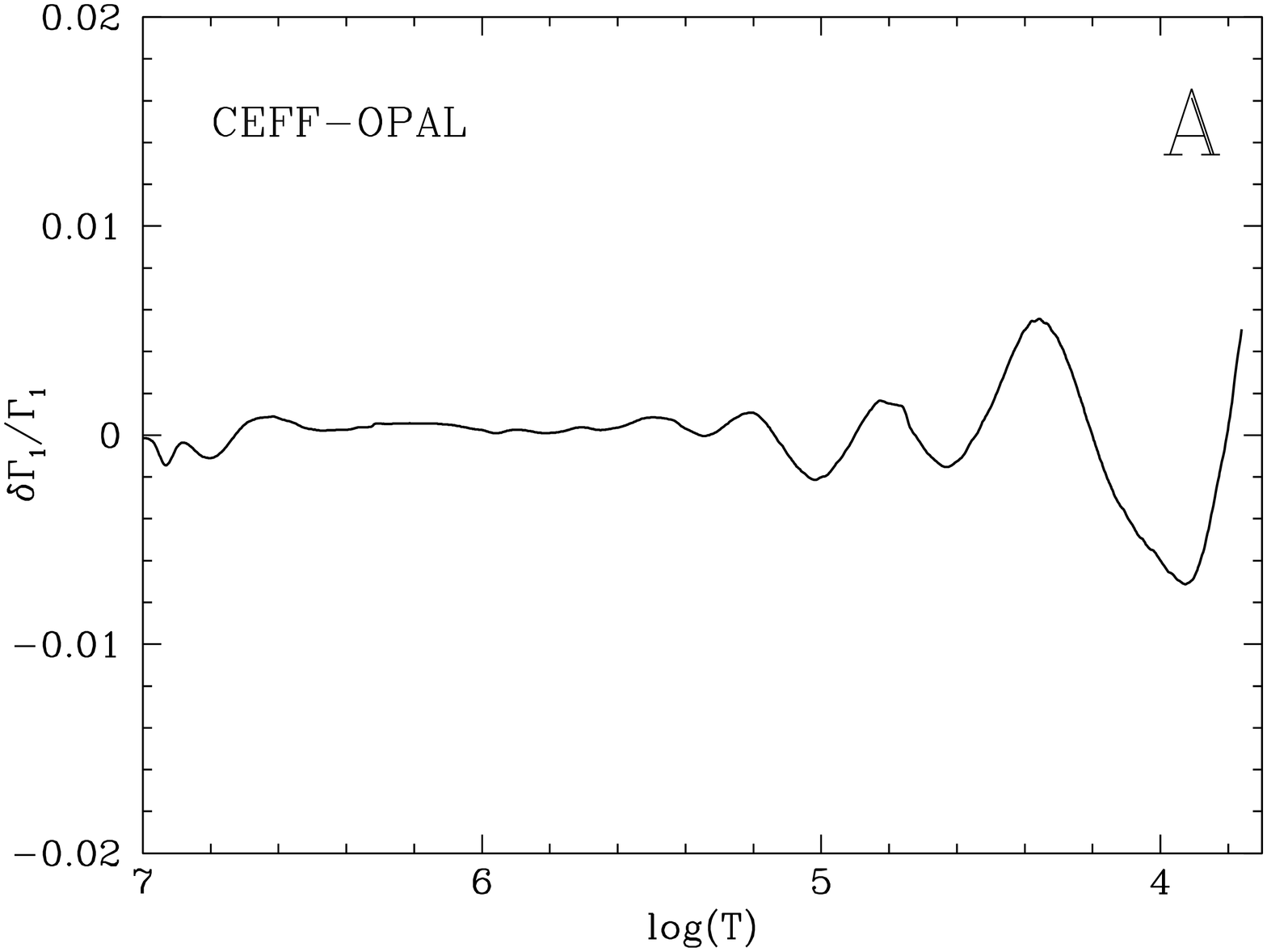}
  \includegraphics[angle=-90,width=.5\textwidth]{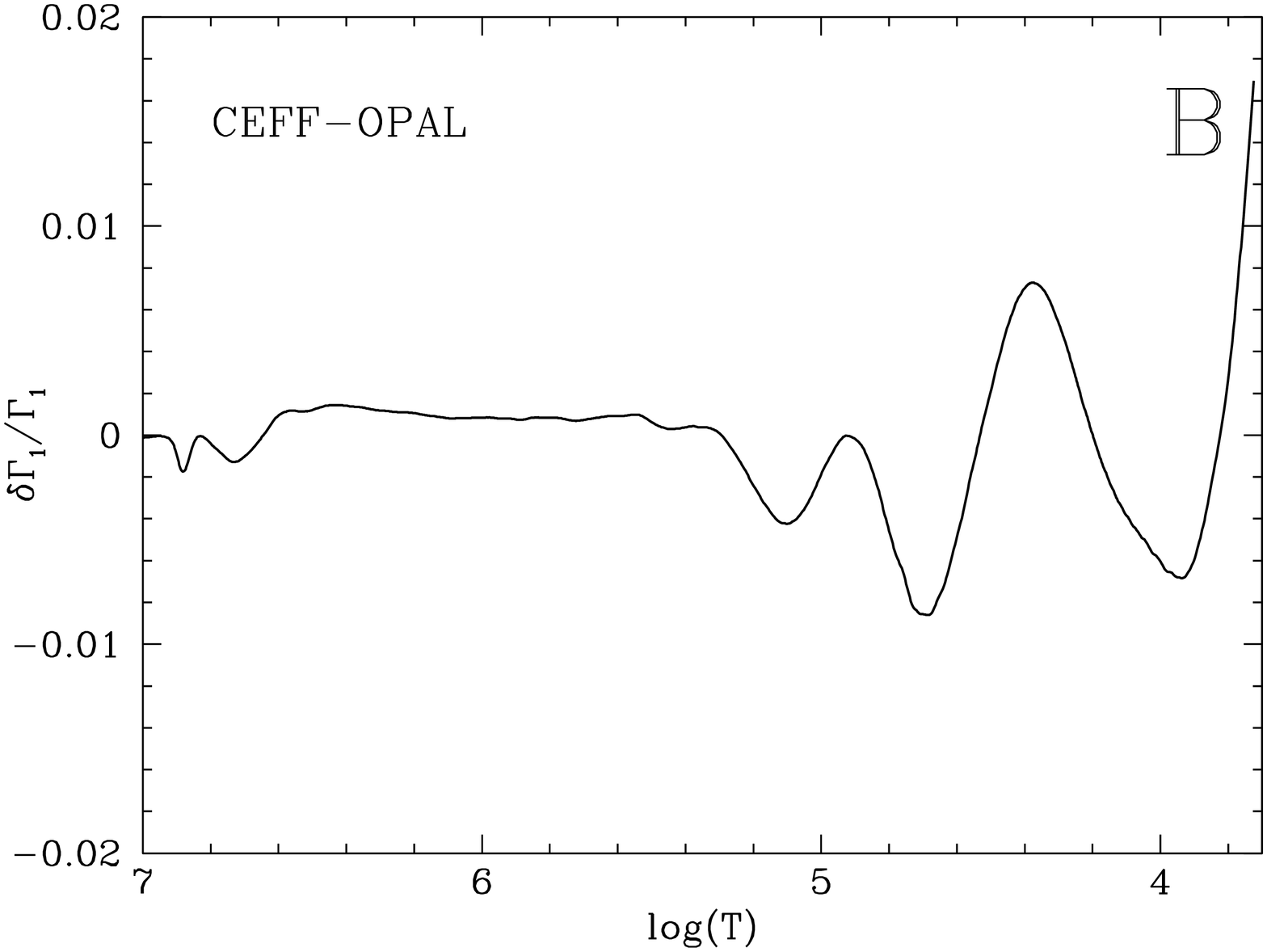}
  \caption{Intrinsic difference in $\Gamma_1$ in model A0 (left panel) and B0 (right panel).}\label{fig:dgamma}
\end{figure}
\begin{figure}
  \includegraphics[angle=-90,width=.5\textwidth]{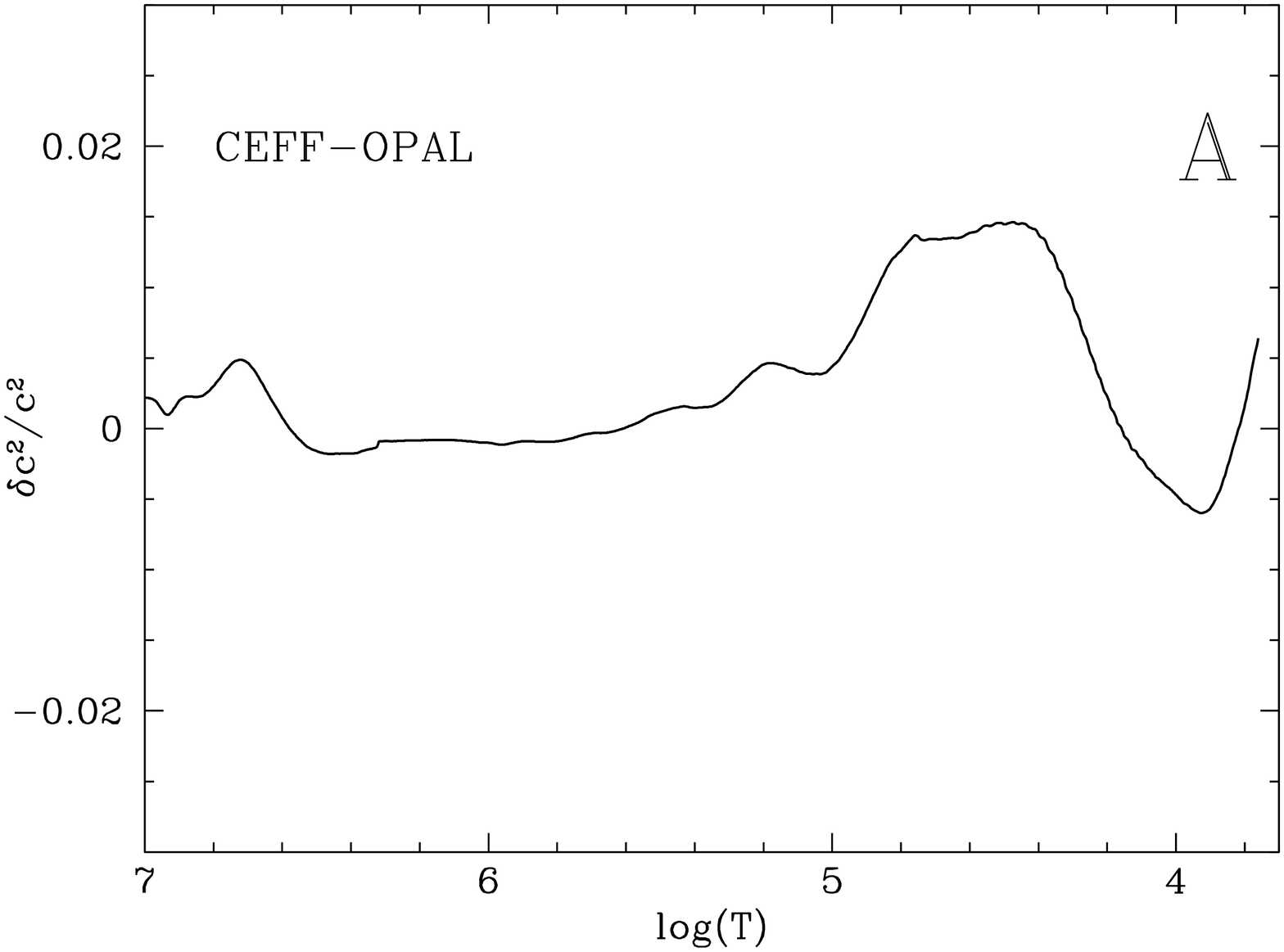}
  \includegraphics[angle=-90,width=.5\textwidth]{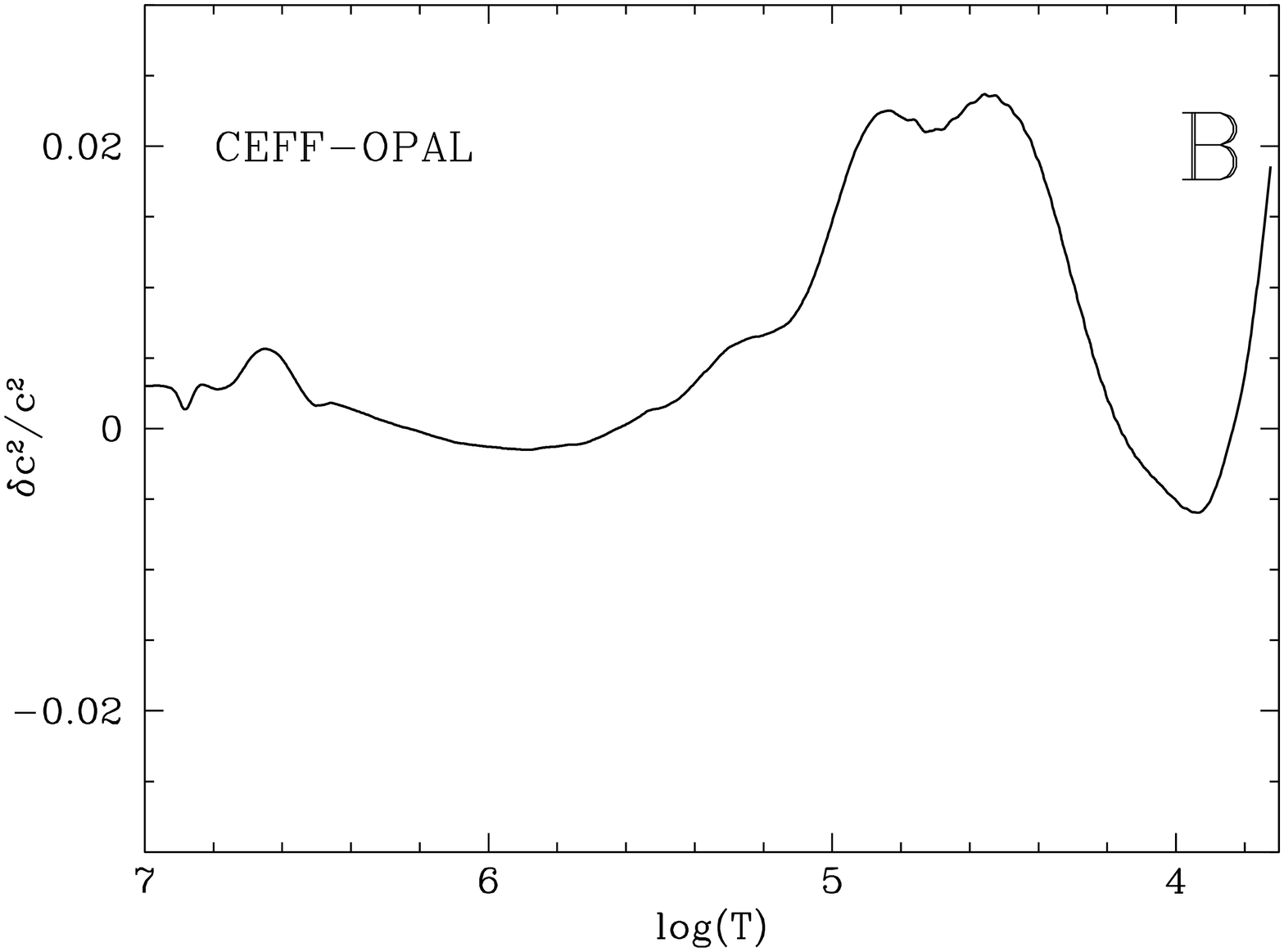}\label{fig:dc2}
  \caption{Intrinsic difference in squared adiabatic sound speed $c^2$.}
\end{figure} 
\end{center}

\section{Calibration using OPAL and CEFF}
\begin{center}
\begin{figure}
  \includegraphics[angle=-90,width=.5\textwidth]{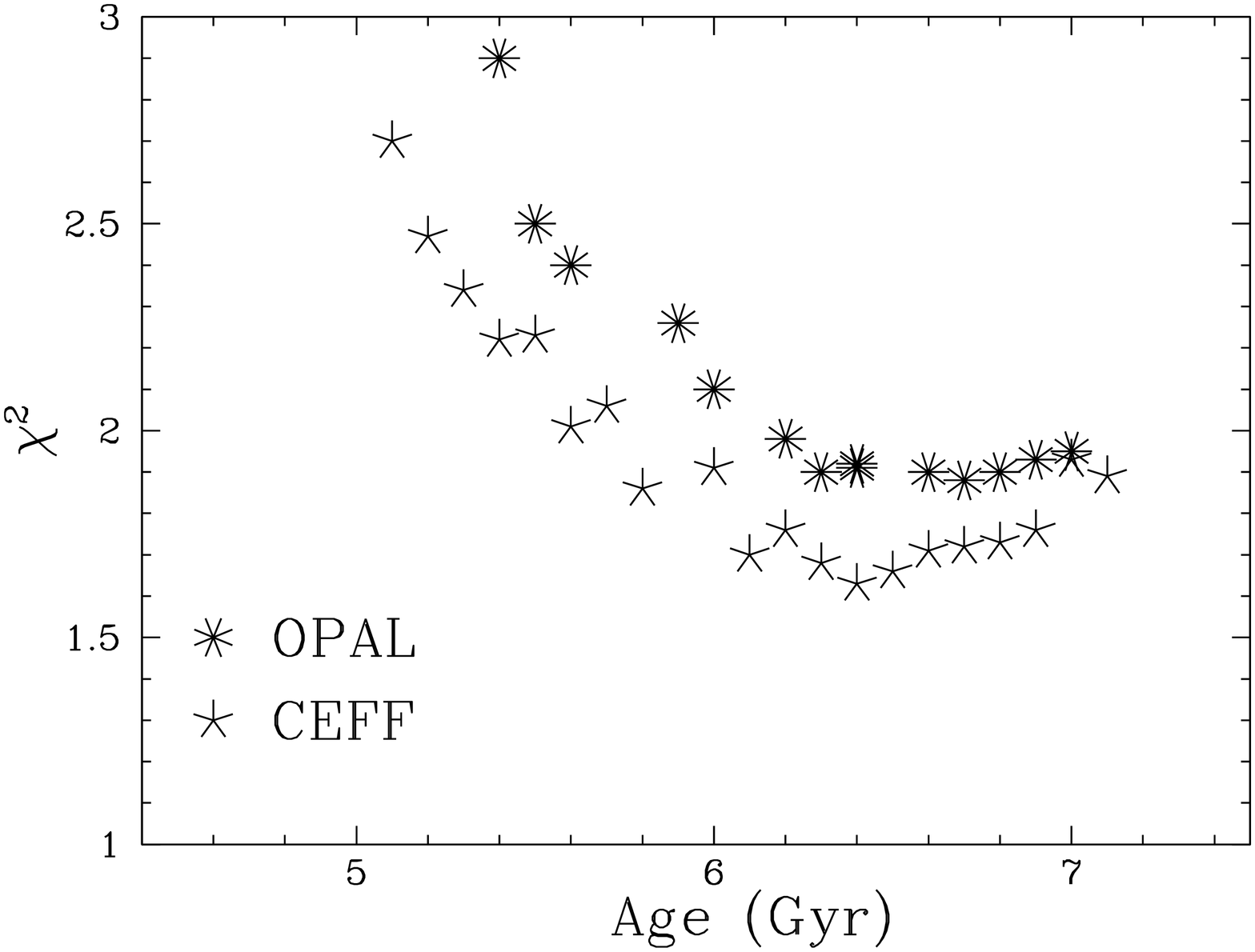}
  \includegraphics[angle=-90,width=.5\textwidth]{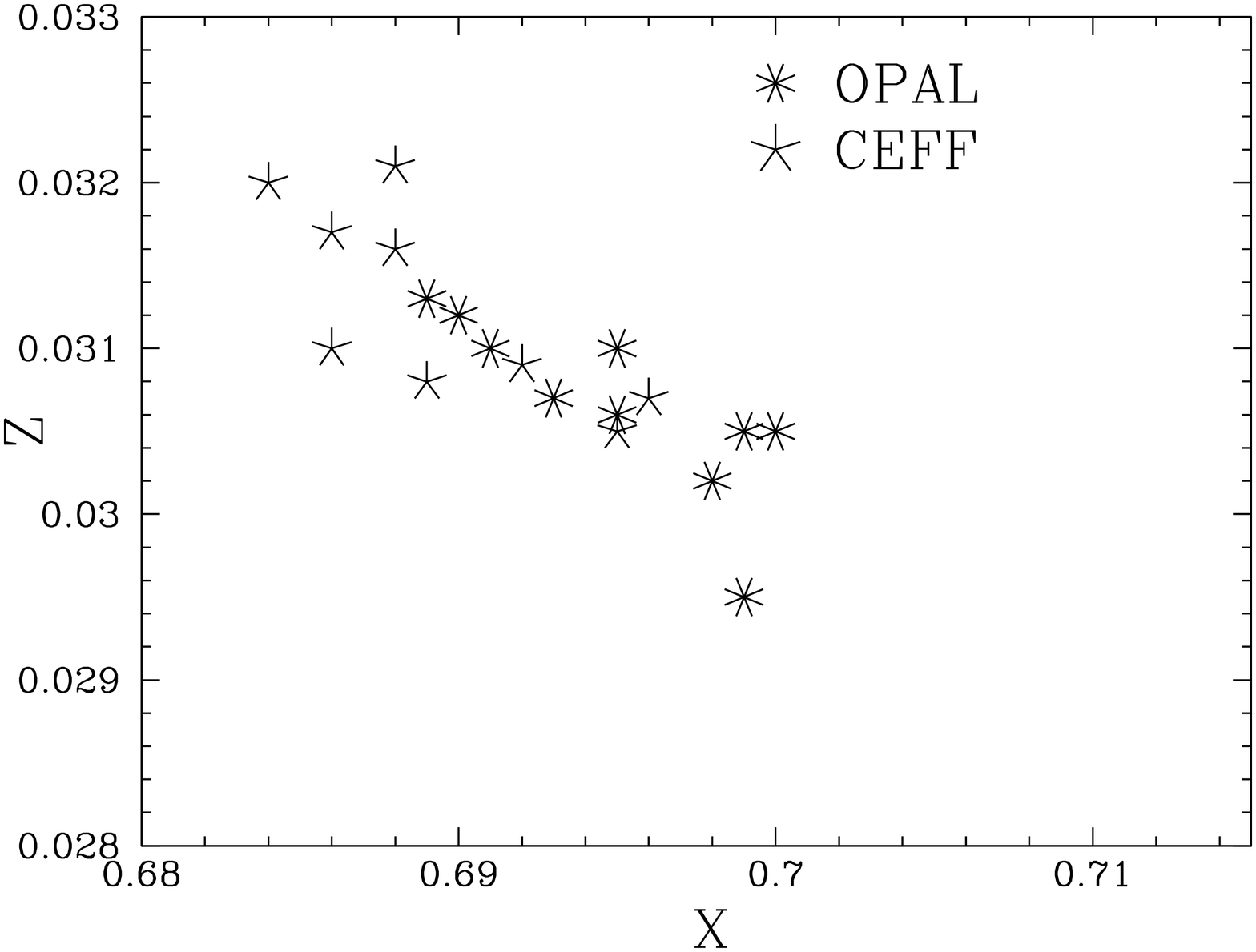}\label{fig:calib}
  \caption{$\chi^2$ of the best model as a function of age. Initial chemical composition of models calibrated with CEFF and OPAL that have a similar $\chi^2$.}
\end{figure} 
\end{center}

Both CEFF and OPAL lead to a quantitatively ($\chi^2$) similar calibration, nonetheless the difference between model parameters (in particular X,Z and Age) corresponding to good-fit models computed with CEFF and OPAL could represent a useful estimate of systematic uncertainties in the parameters due to the use of a different EOS in the modelling. 
For this purpose we have run our calibration algorithm with both EOS; the results are shown in Fig. \ref{fig:calib}. The initial  helium content and metallicity of models that have a similar fit to the observational constraints slightly differ (right panel) and no significant difference in the age estimated using CEFF and OPAL calibrations is found (left panel).
\section{Conclusions}
In agreement with previous works (e.g. \cite{brown94}) we find that present day accuracy of observed  p-mode oscillations in solar-type stars is not sufficient to isolate and detect any ``equation of state effects'' in the calibration of the binary system $\alpha$ Centauri. 
We show nonetheless that the comparison between calibrations performed using different equations of state could provide an additional source of systematic uncertainty on the model parameters.


\begin{theacknowledgments}
The author acknowledges financial support from the Prodex-ESA Contract 15448/01/NL/Sfe(IC) and is thankful to J. Montalb\'an, A. Noels and R. Scuflaire for very useful discussions and help in the redaction of the paper. Thanks are also due to M.P. Di Mauro and to the organizers of the meeting.
\end{theacknowledgments}


\bibliographystyle{aipproc}   


\end{document}